\documentclass[preprint,12pt]{elsarticle}

\usepackage{amssymb}
\usepackage{amsmath}
\usepackage{booktabs}
\usepackage{cite}
\usepackage{hyperref}

\journal{Under Review}

\begin{document}

\begin{frontmatter}



\title{Comparative Analysis of Diffusion Generative Models in Computational Pathology}

\author{Denisha Thakkar\textsuperscript{a}, Vincent Quoc-Huy Trinh\textsuperscript{b}, Sonal Varma\textsuperscript{c}, \\Samira Ebrahimi Kahou\textsuperscript{d,e}, Hassan Rivaz\textsuperscript{a}, Mahdi S. Hosseini\textsuperscript{a,e}}

\affiliation[label1]{
organization={Department of Computer Science and Software Engineering (CSSE), Concordia University}, 
addressline={Montreal}, 
city={QC}, 
postcode={H3H 2R9}, 
state={Quebec}, 
country={Canada}
}

\affiliation[label2]{
organization={University of Montreal Hospital Center}, 
addressline={Montreal}, 
city={QC}, 
postcode={H2X 0C2}, 
state={Quebec}, 
country={Canada}
}

\affiliation[label3]{
organization={Queen's University and Kingston Health Sciences Center}, 
addressline={Kingston}, 
city={ON}, 
postcode={K7L 2V7}, 
state={Ontario}, 
country={Canada}
}

\affiliation[label4]{
organization={Department of Electrical and Software Engineering (ESE), University of Calgary}, 
addressline={Calgary}, 
city={AB}, 
postcode={T2N 1N4}, 
state={Alberta}, 
country={Canada}
}

\affiliation[label5]{
organization={Mila--Quebec Artificial Intelligence Institute}, 
addressline={6666, St-Urbain, \#200}, 
city={Montreal}, 
postcode={H2S 3H1}, 
state={Quebec}, 
country={Canada}
}

\begin{abstract}
Diffusion Generative Models (DGM) have rapidly surfaced as emerging topics in the field of computer vision, garnering significant interest across a wide array of deep learning applications. Despite their high computational demand, these models are extensively utilized for their superior sample quality and robust mode coverage. While research in diffusion generative models is advancing, exploration within the domain of computational pathology and its large-scale datasets has been comparatively gradual. Bridging the gap between the high-quality generation capabilities of Diffusion Generative Models and the intricate nature of pathology data, this paper presents an in-depth comparative analysis of diffusion methods applied to a pathology dataset. Our analysis extends to datasets with varying Fields of View (FOV), revealing that DGMs are highly effective in producing high-quality synthetic data. An ablative study is also conducted, followed by a detailed discussion on the impact of various methods on the synthesized histopathology images. One striking observation from our experiments is how the adjustment of image size during data generation can simulate varying fields of view. These findings underscore the potential of DGMs to enhance the quality and diversity of synthetic pathology data, especially when used with real data, ultimately increasing accuracy of deep learning models in histopathology. Code is available from \href{https://github.com/AtlasAnalyticsLab/Diffusion4Path}{https://github.com/AtlasAnalyticsLab/Diffusion4Path}.
\end{abstract}

%


\end{frontmatter}


\section{Introduction}
Histopathology involves diagnosing diseases by closely inspecting gigapixel tissue slides of microscopic structures to identify their characteristics \citep{jahn2020digital}. Computational pathology has gained significant momentum, bringing about a transformative change in the field of cancer diagnostics since the digitization of pathology slides \citep{baxi2022digital}. In addition, the success of deep learning methods \citep{janowczyk2016deep} has led to the development of numerous models that enhance histopathology diagnosis and can also assist pathologists in their diagnosis workflow \citep{van2021deep, echle2021deep}. However, these models require large volumes of both annotated or unannotated data for effective analysis and cancer screening solutions \citep{cho2015much}.

Two primary challenges, related to data issues, influence the development and implementation of computational pathology foundation models in clinical settings.
First, privacy concerns severely limit the availability of data, making many pathological data inaccessible to the public \citep{price2019privacy}. Patient confidentiality and strict regulatory requirements often prevent the sharing of medical data, which is crucial for training and validating deep learning models. This lack of accessible data limits the ability to create the most-optimized and generalizable models. Therefore, without access to diverse data for research, it becomes challenging to develop models that are truly representative and effective across various patient populations and conditions.

Secondly, the limited quantity and variable quality of publicly available pathology data significantly undermine the performance of foundation models, especially for common lesions \citep{10341042}. Public datasets often lack the breadth and depth needed to train comprehensive models. The available data can be inconsistent in terms of resolution, staining techniques, and annotation quality \citep{hosseini2024computational}. Without sufficient high-quality data, models may struggle to accurately diagnose and improve learning for these conditions.

Another major challenge with pathology data is its complexity, which requires experienced pathologists to analyze and interpret it. It deals with gigapixel images that have multiple magnification levels and unique features \citep{kuklyte2021evaluation}, which are different from typical computer vision data. Studying these images and creating synthetic datasets requires a deep understanding of their unique traits. This knowledge is essential to accurately replicate pathological data in the synthesis process, helping to advance data processing and research. These challenges highlight the need for innovative solutions to improve data sharing and improve the quality of publicly available pathology datasets. Addressing these issues is essential for advancing the field of computational pathology and improving diagnostic workflow capabilities in clinical settings.

Generative models in pathology are still in their early stages but they hold immense potential, especially in the realm of synthetic medical image generation, which can help address the challenges mentioned above in this field. Moreover, by creating synthetic images, generative models can become helpful in discovering new patterns and regularities that provide deeper insights for rare tissue types \citep{chen2021synthetic}. In particular of Generative models, Generative Adversarial Networks (GANs) \citep{GANs2014Goodfellow} have set a high standard in creating high-fidelity and quality data patterns, greatly benefiting fields that demand high realism, such as medical imaging \citep{tang2021disentangled, zhou2020hi, quiros2019pathologygan, kapil2018deep, falahkheirkhah2023deepfake}. Despite GANs' achievements, their real application has many challenges, such as mode collapse and training instability, which are critical setbacks in sensitive applications such as medical imaging. 

Recently, diffusion generative models (DGMs) \citep{ho2020denoising,song2019generative,song2020score,dhariwal2021diffusion, nichol2021improved,song2020improved} have emerged with superior image generation capabilities and great model stability.  To date, diffusion models have been found to be useful in a wide variety of areas, ranging from generative modeling tasks such as image generation \citep{dhariwal2021diffusion}, image super-resolution  \citep{li2022srdiff} to discriminative tasks such as image segmentation \citep{oh2023diffmix}, classification  \citep{han2022card}. DGMs offer a structured approach based on a strong mathematical foundation, which makes them more reliable to use and offers great flexibility in terms of image generation.  
Although DGMs have been applied in various fields for multiple tasks, such as text-conditioned generation \citep{yellapragada2024pathldm},  image generation \citep{moghadam2023morphology}, and large-scale image generation \citep{graikos2024learned}, their potential for generating synthetic datasets and their applicability in tissue classification are still less explored. 

\begin{figure}[htp]
    \centering
    \includegraphics[width=1\linewidth]{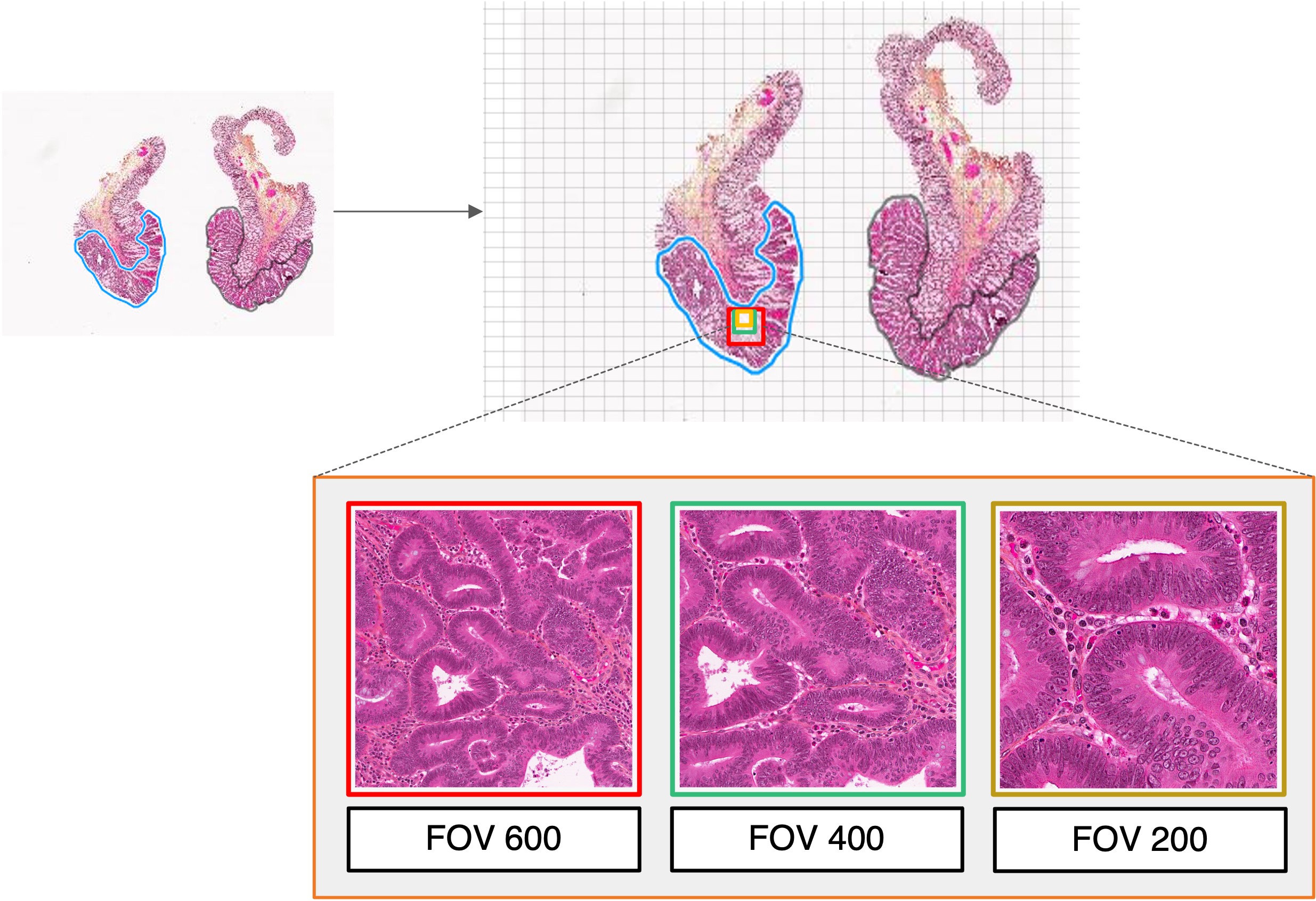}
    \caption{The process of selecting and magnifying different regions from a histopathology slide to obtain patches with varying Fields of View (FOV). The images demonstrate the detail captured at FOV 600, FOV 400, and FOV 200, representing the data's range of granularity and the model's ability to maintain clarity at different magnification levels.}
    \label{fig:fov_overview}
\end{figure}

In our research, we have addressed this gap using extracted datasets for different image resolution from the WSI dataset. By doing so, we were able to compare the differences in synthesizing various Field-Of-View (FOV)  \citep{basavanhally2011boosted} from the same whole slides, specifically focusing on FOVs of 224 and 336. Different regions of a histopathology slide may require different levels of detail, from high-level overviews to detailed cellular structures as demonstrated in Figure \ref{fig:fov_overview}.  We also provide a novel study on the effects of prompting patch size, which further helps to understand how different patch sizes can influence FOV detail generation. This study shows that by prompting with different sizes, DGMs can generate images with varying levels of detail that are unseen during training. We also show how these synthetic datasets match robustness and accuracy of deep learning classifiers as well.

This study represents a novel exploration of DGMs in the field of computational pathology. It offers a comprehensive comparison of various baseline methods and showcases the ability of diffusion models to create coarse features and generate images with varying patch size.

The contributions of this paper are as follows:
\begin{itemize}
\item \textbf{Comprehensive Analysis:}
This paper includes a comprehensive study  that compares various diffusion generative methods. The study provides a detailed analysis of the strengths and weaknesses of different techniques, offering valuable information on their performance and applicability in medical imaging. Research contributes to a better understanding of how different generative models perform in the context of pathology.
\item \textbf{Novel Study of Diffusion Models in Pathology:}
The publication not only uses baseline methods to synthesize images but also explores the adaptability of diffusion models by generating images of varying patch sizes. This novel study explores the innovative capabilities of diffusion models, particularly their unique ability to generate images with varying FOVs by prompting them with different patch sizes. It highlights how these models can produce high-quality synthetic pathology images that accurately reflect different features at different levels. By generating images with varying FOVs, diffusion models can capture the intricate details and structural variations of tissue samples. This capability can become useful to enhance the versatility and effectiveness of models in research and educational settings.
\item \textbf{Comparison of FOVs:}
In our investigation, we have addressed critical gaps utilizing multiple subset datasets from singe dataset. By synthesizing various FOVs from the same dataset, specifically focusing on FOVs of 224 and 336, we have been able to compare the differences in FID score by diffusion-generative models (DGM).
\item \textbf{Exploration of Synthetic Dataset Applicability:}
This research also explores the applicability of synthetic datasets generated by diffusion models in enhancing deep learning classifiers. The study evaluates how well these synthetic images can be used to train and improve the performance of deep learning models, specifically in the context of medical image classification. One of the key contributions of this paper is the demonstration of improved classification performance by incorporating synthetic datasets alongside real images. Empirical evidence is presented, showing that the addition of high-quality synthetic images can significantly boost the performance of classifiers, leading to a more robust classifier.
\end{itemize}

\section{Review of Diffusion Generative Methods}
\begin{figure*}[ht]
        \centering
        \includegraphics[width=1\textwidth]{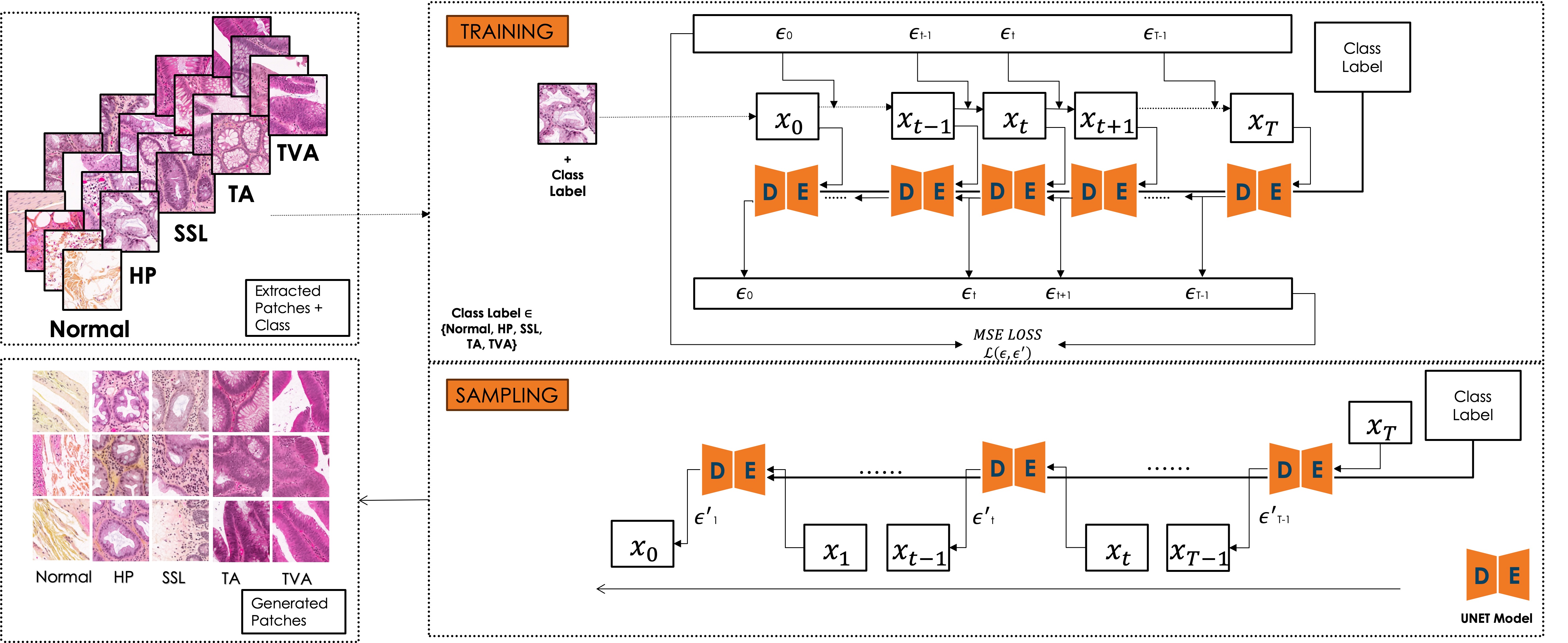}
        \caption{Diffusion Generative Models framework used in the pathology dataset: The Training phase (top-right) details the data points' progression from \( x_0 \) to \( x_T \) with noise levels \( \epsilon_0 \) to \( \epsilon_T \), where the model adds noise to the original data and estimates the reverse process guided by class information to understand tissue details. The Sampling phase (bottom-right) reverses the diffusion from a noisy state \( x_T \) to the original data point \( x_0 \), by iteratively denoising using noise predictions \( \epsilon' \) and learned parameters, and prompting the class label to generate images of specific tissue types.}
        \label{fig:diffusion_process}
    \end{figure*}

Diffusion models \citep{sohl2015deep,ho2020denoising,song2019generative} are a powerful class of probabilistic generative models that are used to learn complex data distributions. The purpose of this framework is also to demonstrate how various methods can be integrated seamlessly. We explain each category and method in a manner that allows any standalone model or technique to be easily plugged into the framework. The framework is explained as follows:

\begin{enumerate}

\item \textbf{Training:}
  \begin{itemize}
    \item \textbf{Forward Process:} Gaussian noise is introduced into the input images, with the magnitude determined randomly by a scheduler, resulting in varying levels of noise. This allows the model to explore a range of image noises, from subtle distortions to isotropic Gaussian noise. The noisy image, along with embedding vectors representing the noise level and class labels, is fed into a U-Net-based neural network.
    \item \textbf{Reverse Process:} The U-Net predicts the noise added to the images. This prediction is compared to the actual noise to calculate the loss, guiding the model's training process.
  \end{itemize}

\item \textbf{Sampling:}
  The process starts with pure isotropic noise. The trained network predicts the corresponding noise level, and the difference between the predicted noise and the initial noise creates a less noisy image. This image is iteratively refined through multiple steps to generate the final image.

This whole process is explained well in the figure \ref{fig:diffusion_process}.
\end{enumerate}

\subsection{Denoising Diffusion Probabilistic Models}
\label{subsec:ddpm}

Denoising Diffusion Probabilistic Models (DDPM) \citep{ho2020denoising} follow the same training and sampling methods explained above, with both the forward and reverse processes occurring in the pixel space. In the forward process, Gaussian noise is added to the image \(x_0\) at each timestep \(t\) according to:
\begin{equation}
q(\mathbf{x}_t | \mathbf{x}_{t-1}) = \mathcal{N}(\mathbf{x}_t; \sqrt{1 - \beta_t} \mathbf{x}_{t-1}, \beta_t \mathbf{I})
\end{equation}
where \(\beta_t\) is the variance schedule. The reverse process aims to denoise the image step-by-step using a learned model \(p_\theta\), represented as:
\begin{equation}
p_\theta(\mathbf{x}_{t-1} | \mathbf{x}_t) = \mathcal{N}(\mathbf{x}_{t-1}; \boldsymbol{\mu}_\theta(\mathbf{x}_t, t), \boldsymbol{\Sigma}_\theta(\mathbf{x}_t, t))
\end{equation}
The objective function to train the model is:
\begin{equation} \label{eqn:ddpm_loss}
L_{t}^\text{simple} = \mathbb{E}_{t \sim [1, T], \mathbf{x}, \boldsymbol{\epsilon}_t} \Big[\|\boldsymbol{\epsilon}_t - \boldsymbol{\epsilon}_\theta(\mathbf{x}_t, t)\|^2_2 \Big]
\end{equation}

\subsection{Latent Diffusion Models}
\label{subsec:ldm}
Latent Diffusion Models (LDM) \citep{rombach2022high} operate in a lower-dimensional latent space, which is learned by an autoencoder. The training and sampling stages follow the same principles as the general diffusion models described above, but with processes occurring in the latent space. An extra layer is added to encode images into this latent space before the diffusion processes begin. The forward process is defined as:
\begin{equation}
q(\mathbf{z}_t | \mathbf{z}_{t-1}) = \mathcal{N}(\mathbf{z}_t; \sqrt{1 - \beta_t} \mathbf{z}_{t-1}, \beta_t \mathbf{I})
\end{equation}
where \(\mathbf{z}_t\) represents the latent variables. The reverse process denoises the latent variables step-by-step:
\begin{equation}
p_\theta(\mathbf{z}_{t-1} | \mathbf{z}_t) = \mathcal{N}(\mathbf{z}_{t-1}; \boldsymbol{\mu}_\theta(\mathbf{z}_t, t), \boldsymbol{\Sigma}_\theta(\mathbf{z}_t, t))
\end{equation}
LDMs also incorporate cross-attention mechanisms within the architecture, enhancing conditional image synthesis. After the latent space processing, the autoencoder transforms the latent variables back into images. The objective function for training remains similar:
\begin{equation} \label{eqn:ldm_loss}
L_{t}^\text{simple} = \mathbb{E}_{t \sim [1, T], \mathbf{z}, \boldsymbol{\epsilon}_t} \Big[\|\boldsymbol{\epsilon}_t - \boldsymbol{\epsilon}_\theta(\mathbf{z}_t, t)\|^2_2 \Big]
\end{equation}

\subsection{Conditioning}
\label{subSec:conditioning}
Recent advancements in DPMs have introduced class-conditional generation, where additional class-related information is incorporated to guide the generation process. Our findings revealed that class-conditional generation significantly enhances the fidelity of the generated images to specific class characteristics. As the guidance scale increased, the generated images exhibited more precise and accurate representations of the target classes.

\label{subsubsec:cfg}
Classifier-Free Guidance (CFG) \citep{ho2022classifier} is a technique that enables the generation of high-quality samples without relying on a classifier, addressing the limitations associated with classifier guidance. CFG modifies the score function in a way that emulates the effects of classifier guidance, but without using an explicit classifier. The approach involves training an unconditional denoising diffusion model alongside the conditional model, using a single neural network to parameterize both. The sampling process utilizes a combination of the conditional and unconditional score estimates, allowing for effective guidance without a classifier. This results in the production of high-quality synthetic images that are both varied and representative of the original dataset, enhancing the model's performance in generating realistic images.

\begin{equation}
\epsilon_t = (1 + w) * \epsilon_\theta(x_t, c) - w *\epsilon_\theta(x_t)
\end{equation}

Here, $\epsilon_\theta(x_t, c)$ is conditional model and $\epsilon_\theta(x_t)$ is unconditional model. $w$ is used as a guidance scale.
\subsection{More Sampling Choices}
\label{subsec:sampling}
To explore different sampling methods and their effects on image generation and model performance, we incorporated two additional techniques: DDIM and Epsilon Scaling. DDIM accelerates image generation by introducing a non-Markovian process that redefines the diffusion process, utilizing a subset sampling strategy for faster sampling without compromising model performance. The reverse diffusion process in DDIM is defined as:
\begin{equation}
\mathbf{x}_{t-1} = \sqrt{\alpha_{t-1}} \left( \frac{\mathbf{x}_t - \sqrt{1 - \alpha_t} \epsilon_\theta(\mathbf{x}_t)}{\sqrt{\alpha_t}} \right) + \sqrt{1 - \alpha_{t-1} - \sigma^2_t} \epsilon_\theta(\mathbf{x}_t) + \sigma_t \epsilon_t
\end{equation}
where the variance \(\sigma_t\) is given by:
\begin{equation}
\sigma_t = \eta \sqrt{\frac{1 - \bar{\alpha}_{t-1}}{1 - \bar{\alpha}_t} \cdot \beta_t} = \eta \sqrt{\tilde{\beta}_t}
\end{equation}
In DDIM, setting \(\eta = 0\) eliminates noise, making it equivalent to DDPM, while \(\eta = 1\) maintains the standard diffusion model, allowing interpolation between DDIM and DDPM. To address exposure bias, we employed Epsilon Scaling \citep{ning2023elucidating}, which scales the epsilon value using a linear function, ensuring consistency between training and sampling, thereby reducing bias and improving sample quality. The epsilon value is scaled as:
\begin{equation}
\epsilon_t = \frac{\epsilon_{\theta}(\mathbf{x}_t, t)}{\lambda_t}
\label{eq:designlambda}
\end{equation}
where \(\lambda_t = \lambda_t k + b\). The scaled \(\epsilon_\theta\) is then used in the DDPM equation:
\begin{equation}
\mathbf{x}_{t-1} = \frac{1}{\sqrt{\alpha_t}} \left( \mathbf{x}_t - \frac{1 - \alpha_t}{\sqrt{1 - \bar{\alpha}_t}} \boldsymbol{\epsilon}_t \right) + \sigma_t z
\end{equation}

\section{Experiment Details}
\subsection{KGH Dataset}\label{sec:kgh-dataset}
KGH (Kingston General Hospital) datasets is made of \textbf{1037} WSI of healthy colon and four types of colonic polyps. The pathological slides are annotated on the Region of Interest (ROI) level. The WSIs are rectangular and of different sizes but their size is around 60k $\times$ 100k pixels and they take approximately 1 GB of memory. Because of the size of the WSIs and our computational resouces, we cannot process the slides as-is. Therefore, we have to extract some tiles of these images, called patches. To extract patches, we have to precise the FOV of the patches. We have used FOV as 224 from magnification 20X. We resized these patches to 128X128. 
\newline KGH datasets presents healthy colon tissue and four different types of colon polyps: Tubular Adenoma (TA), Sessile Serrated Lesions (SSL), Hyperplastic Polyps (HP), Tubulovillous Adenoma (TVA). HP is a non-neoplastic polyp; whereas the other 3 types SSA, TA and TVA have clonal abnormality at the molecular levels, and hence are considered neoplastic. All the 3 polyps have the potential to progress to adenocarcinoma.

\begin{figure}[ht]
    \centering
    \includegraphics[width=1\linewidth]{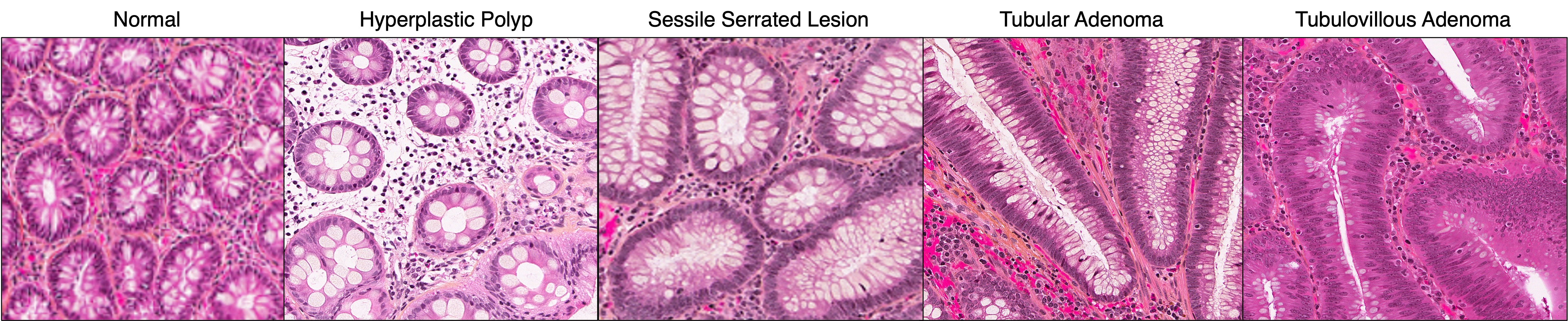}
    \caption{Samples from Cancer Tissue and Normal WSI}
    \label{fig:classes}
\end{figure}

We can visualize patches from these classes in Figure \ref{fig:classes}. Despite some visual differentiation in certain regions among these classes, distinguishing them can be challenging without a pathologist's expertise. Understanding these classes is crucial for making accurate analogies with our generated images. The visual distinctions, although subtle, provide necessary context for training our models. To manage the high-dimensional nature of these pathology images, we extract smaller sections referred to as patches. To extract these patches, we specify the FOV. We also eliminated noisy images from the original set.

Patches were extracted from the annotated WSIs at specified FOVs. For our experiments, we maintained a consistent patch resolution of 1.75. This approach allowed us to observe how different FOVs provide various perspectives of the same tissue at the same magnification. Specifically, we used FOVs of 224 and 336 for creating the PKGH dataset. These FOVs were selected to ensure that the patches offered a comprehensive view of the cellular and tissue structures necessary for training generative models. The chosen FOVs are not excessively large to avoid having a limited number of patches available for training and classification.
 
\subsection{The Architecture}
    
 We used a UNet model based on \citep{dhariwal2021diffusion} for our experiments. We converted 128 $\times$ 128 images to noise, which were then fed into the UNet network. The network predicted the amount of noise added to these images and outputted the noise prediction. We employed a loss function that compares the predicted noise with the actual noise from the stochastic process. In Diffusion Models, Gaussian noise is iteratively added to the original image according to a variance schedule with a large total number of steps (T = 1000). The UNet model also incorporates  timestep and class embeddings to learn class-specific features.
    
We implemented the LDM architecture from \citet{rombach2022high}, which consists of the Variational Autoencoder (VAE), the U-Net denoiser, and added the class embedding. We used an autoencoder based on vector quantization, VQ-autoencoder, referred to as VQF4-DM \citep{van2017neural}.

    \subsection{Evaluation Metrics}   
    A classic approach for generative model's evaluations is to compare the log-likelihoods of models. This approach,
    however, has several shortcomings. A model can achieve high likelihood, but
    low image quality, and conversely, low likelihood and high image quality. The two most common GAN \citep{GANs2014Goodfellow} evaluation measures are Inception Score (IS) \citep{salimans2016improved}
    and Fréchet Inception Distance (FID) \citep{heusel2017gans}.  
    
    Inception Score (IS) rely on a pre-existing classifier
    (InceptionNet) \citep{szegedy2015going} trained on ImageNet. IS computes the
    KL divergence between the conditional class distribution and the marginal class distribution over the generated data.  IS does not capture intra-class diversity, is insensitive to the prior distribution over labels (hence is biased towards ImageNet dataset and Inception model. Therfore, we would not be using it for our dataset as it has completely different data than Imagenet.  However, we would be using FID and KID.
    
    FID \citep{heusel2017gans} calculates the
    Wasserstein-2 (a.k.a Fréchet) distance between multivariate Gaussians fitted to the embedding space of the Inception-v3 network of generated and real images. The Kernel Inception Distance (KID) \citep{binkowski2018demystifying} aims to improve on FID by relaxing the Gaussian assumption. KID measures the squared Maximum Mean Discrepancy (MMD) between the Inception representations of the real and generated samples using a polynomial kernel. This is a non-parametric test so it does not have the strict Gaussian assumption, only assuming that the kernel is a good similarity measure. 

    We further use deep learning based classifiers approach to compare original vs generated data to show how synthetic data is actually applicable in real life. This is separately explained in Section.
    
\subsection{Experiments}
\subsubsection{Comparative analysis}
In our research, we analyze the effects of various Diffusion Generation Models (DGM) on a pathology dataset, focusing on comparisons rather than finding the best method. We examine two models: DDPM and LDM, exploring their performance under pathological data conditions.

LDMs, operating in a learned latent space, are more efficient than pixel-based designs, while DDPMs apply diffusion directly to input images, which is useful for capturing complex patterns in medical images. LDMs generate images with fewer timesteps, making them faster for applications requiring rapid synthesis.

We also utilize classifier guidance to balance mode coverage and sample fidelity in post-training conditional models. This technique, while improving sample quality, may increase computational costs. Both DDPM and LDM are trained on the same pathology datasets with classifier guidance, allowing us to observe and evaluate their visual outputs and subtle improvements, essential for analyzing large pathological images.

\subsubsection{Analysis on different patch size}
In this experiment, we introduced a novel hyperparameter, Patch Size, during the image generation phase to observe its effect on the tissue structure in histopathology images. By varying the patch size, we found that it significantly enhances the model's ability to capture fine details and intricate structures within the generated images, as the model adapts to the provided patch size, effectively learning the resolution of the patches.

  $$
    \text{FOV} = \text{Patch Size} \times \text{Patch Resolution}
    $$

Patch resolution in histopathology images, typically measured in microns per pixel (mpp), determines the level of detail captured relative to the actual tissue size. High patch resolution captures more detailed information, making it ideal for identifying fine morphological features, while lower patch resolution provides a broader view of tissue structures. Understanding and adjusting patch resolution is crucial for optimizing image analysis.

    We validated our findings by first confirming the patch resolution used earlier, where the FOV was 224 and the patch size was 128, resulting in a patch resolution of 1.75 microns per pixel (mpp) according to the formula mentioned above. Using this patch resolution (1.75 mpp) and varying the patch sizes, we calculated the corresponding FOVs:
    
   \begin{table}[h!]
    \centering
    \begin{tabular}{c|c}
    \hline
    \textbf{Patch Size} & \textbf{FOV)} \\
    \hline
    64 & 112 \\
    96 & 168 \\
    128 & 224 (original) \\
    160 & 280 \\
    192 & 336 \\
    224 & 392 \\
    \hline
    \end{tabular}
    \caption{Patch Sizes and Corresponding Fields of View}
    \label{tab:patchsize}
    \end{table}
    
\subsubsection{Evaluation of Synthetic Pathology Dataset}
    The evaluation of generated pathology images is crucial to determine their quality and similarity to real pathology images. Although evaluation metrics provide valuable information about the quality of generated images, their applicability to real datasets remains uncertain. This study aims to analyze the performance of the generated data set on the ResNet-50 architecture \citep{he2016deep} along with the real data set.

\section{Results}

The Fréchet Inception Distance (FID) scores of our DDPM method demonstrate strong performance when compared with similar studies in the field. These comparisons are crucial for highlighting the effectiveness of our approach in generating high-quality images. The Table \ref{table:comparison-other-studies} compares the FID scores of DDPM method with several studies in the literature, as referenced by \citep{pozzi2023generating}. We only consider these among others as these are based on the same DDPM methodology and architecture:

\begin{table}[ht]
\centering
\begin{tabular}{@{}l|l|l@{}}
\toprule
\textbf{Study} & \textbf{Number of Classes} & \textbf{FID} \\ \hline
Current Study (PKGH\_224) & 5 & \textbf{19.08}\\ \hline
Current Study (PKGH\_336) & 5 & \textbf{18.45} \\ \hline
\citep{pozzi2023generating} & 5 & 35.11 \\ \hline
\citep{moghadam2023morphology} & 1 & 20.11 \\ \hline
\end{tabular}
\caption{Comparison of FID scores with existing studies}
\label{table:comparison-other-studies}
\end{table}

When compared to other studies, our DDPM method's FID scores of 19.08 for the PKGH\_224 dataset and 18.45 for the PKGH\_336 dataset are competitive, demonstrating superior performance over \citep{pozzi2023generating} and comparable results with \citep{moghadam2023morphology}. These results highlight the importance of conditioning in DDPM and the significant impact of the FOV used. The findings underscore the value of FOV selection to achieve better outcomes in histopathological image generation.

\subsection{Comparative Analysis}
\label{subsec:comparative}

\begin{figure}[htp]
\centering
\includegraphics[width=0.7\linewidth]{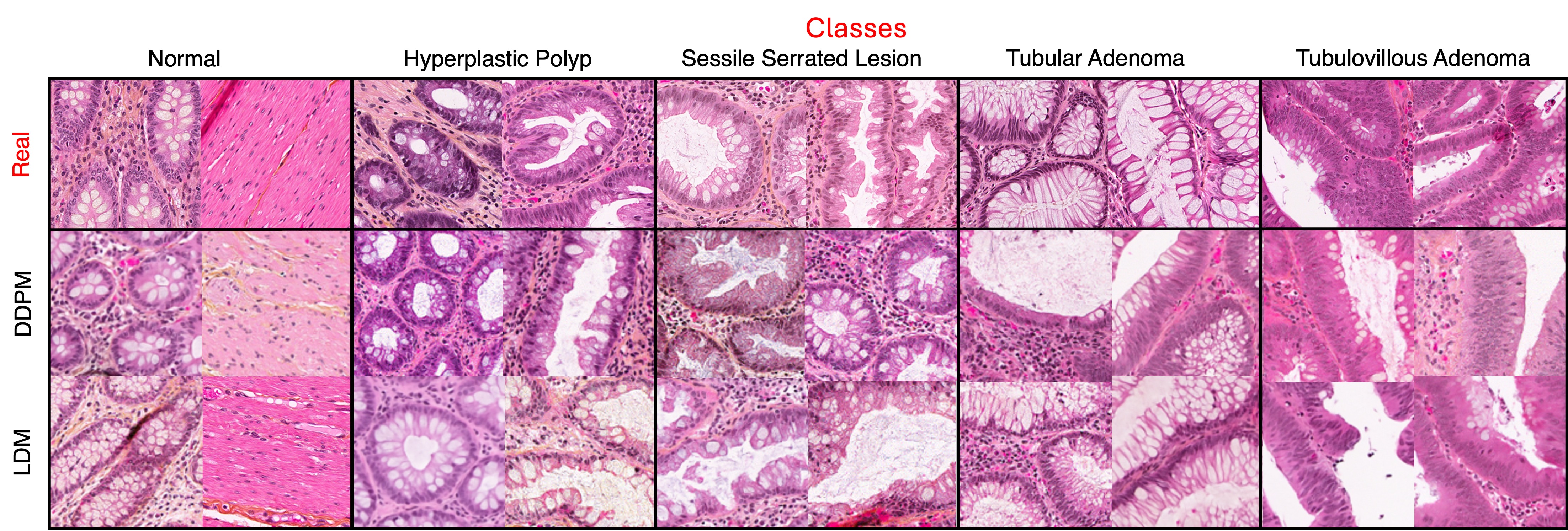}
\caption{Real vs Generated Images from the FOV 224 (DDPM and LDM): The top row shows a representative image from the real dataset for each tissue type, while the bottom row displays a generated image of the same tissue.}
\label{fig:ddpm-ldm-224}
\end{figure}

\begin{figure}[htp]
\centering
\includegraphics[width=1\linewidth]{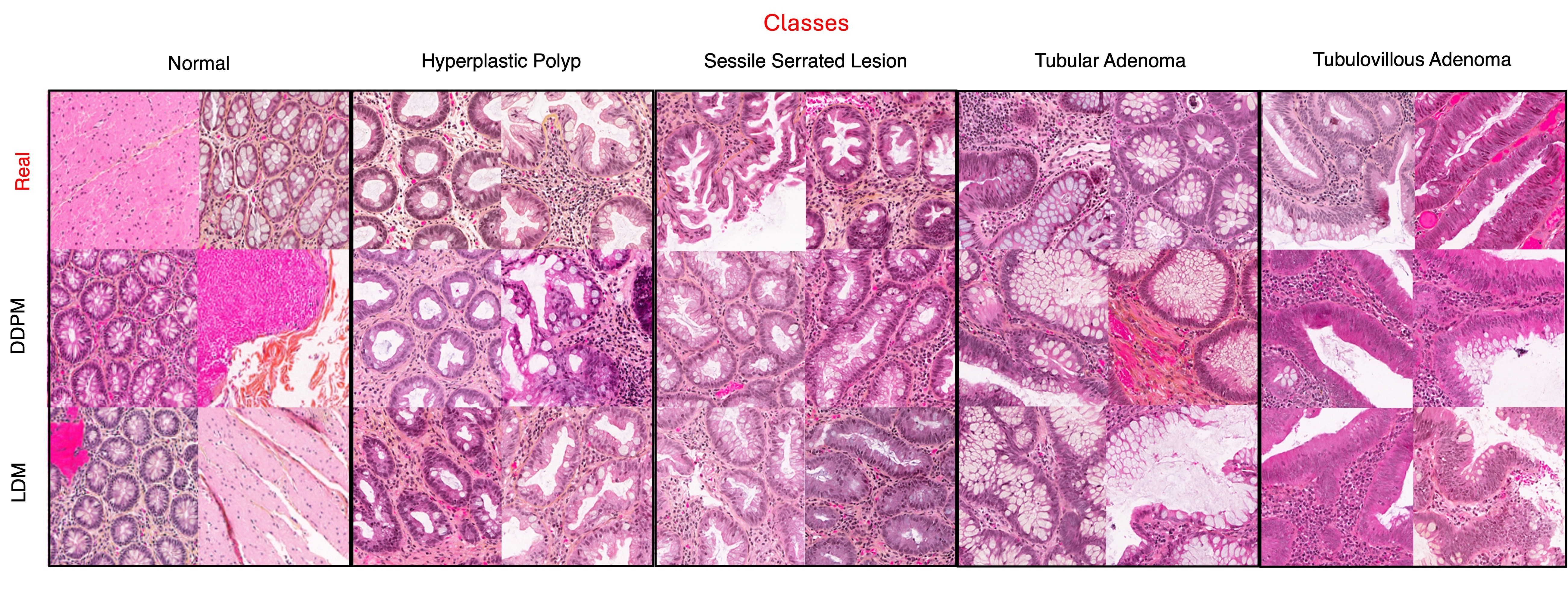}
\caption{Real vs Generated Images from the FOV 336 (DDPM and LDM): The top row shows a representative image from the real dataset for each tissue type, while the bottom row displays a generated image of the same tissue.}
\label{fig:ddpm-ldm-336}
\end{figure}

The analysis focuses on comparing the performance of DDPM and LDM across two datasets: PKGH\_224 and PKGH\_336. 

The Table \ref{table
}    summarizes the results of all methods.

In our study, the PKGH\_224 dataset revealed that the DDPM model, with DDPM sampling, achieved an FID score of 19.08 and a KID score of 0.0134, indicating a strong balance of image quality and diversity. In comparison, the LDM with DDPM sampling produced an FID score of 24.43 and a KID score of 0.0185, showing that while LDM's latent space representation is effective, DDPM slightly outperformed it in this case. When using DDIM sampling, DDPM scored 22.66 (FID) and 0.0154 (KID), while LDM scored 25.56 (FID) and 0.0161 (KID), further demonstrating DDPM's superior performance across sampling methods.

For the larger PKGH\_336 dataset, DDPM again excelled with an FID of 18.45 and a KID of 0.0129 using DDPM sampling, maintaining consistent high performance across different dataset sizes. LDM with DDPM sampling scored 23.10 (FID) and 0.0160 (KID), showing good results but not surpassing DDPM. The DDIM sampling method reinforced this trend, with DDPM scoring 21.34 (FID) and 0.0159 (KID), while LDM scored 26.41 (FID) and 0.0199 (KID).

These results highlight that DDPM consistently outperforms LDM in generating detailed and realistic images, especially in smaller datasets like PKGH\_224. This suggests that DDPM is more effective in controlling pixel space and producing high-quality images, making it a robust choice for synthetic pathology image generation across different dataset sizes.

\begin{table}[ht]
\centering
\begin{tabular}{@{}l|l|l|l@{}}
    \hline
\multicolumn{4}{c}{\textbf{PKGH\_224}} \\     \hline
Training Method & Sampling Method & FID & KID \\     \hline
DDPM & DDPM & 19.08 & 0.0134 \\
& Epsilon Scaling(s=1.014) & 39.88 & 0.0338 \\
& DDIM & 22.66 & 0.0154 \\ \hline
LDM & DDPM & 24.43 & 0.0185\\
& DDIM & 25.56 & 0.0161 \\     \hline
\end{tabular}

\vspace{0.5cm}

\begin{tabular}{@{}l|l|l|l@{}}
    \hline
\multicolumn{4}{c}{\textbf{PKGH\_336}} \\     \hline
Training Method & Sampling Method & FID & KID \\     \hline
DDPM & DDPM & 18.45 & 0.0129 \\
& Epsilon Scaling(s=1.014) & 45.12 & 0.0418 \\
& DDIM & 21.34 & 0.0159 \\ \hline
LDM & DDPM & 23.10 & 0.0160 \\
& DDIM & 26.41 & 0.0199 \\    \hline
\end{tabular}
\caption{FID and KID Scores for Different Models and Sampling Methods on Two Datasets}
\label{table
}
\end{table}

\begin{figure}[ht]
\centering
\includegraphics[width=0.5\textwidth]{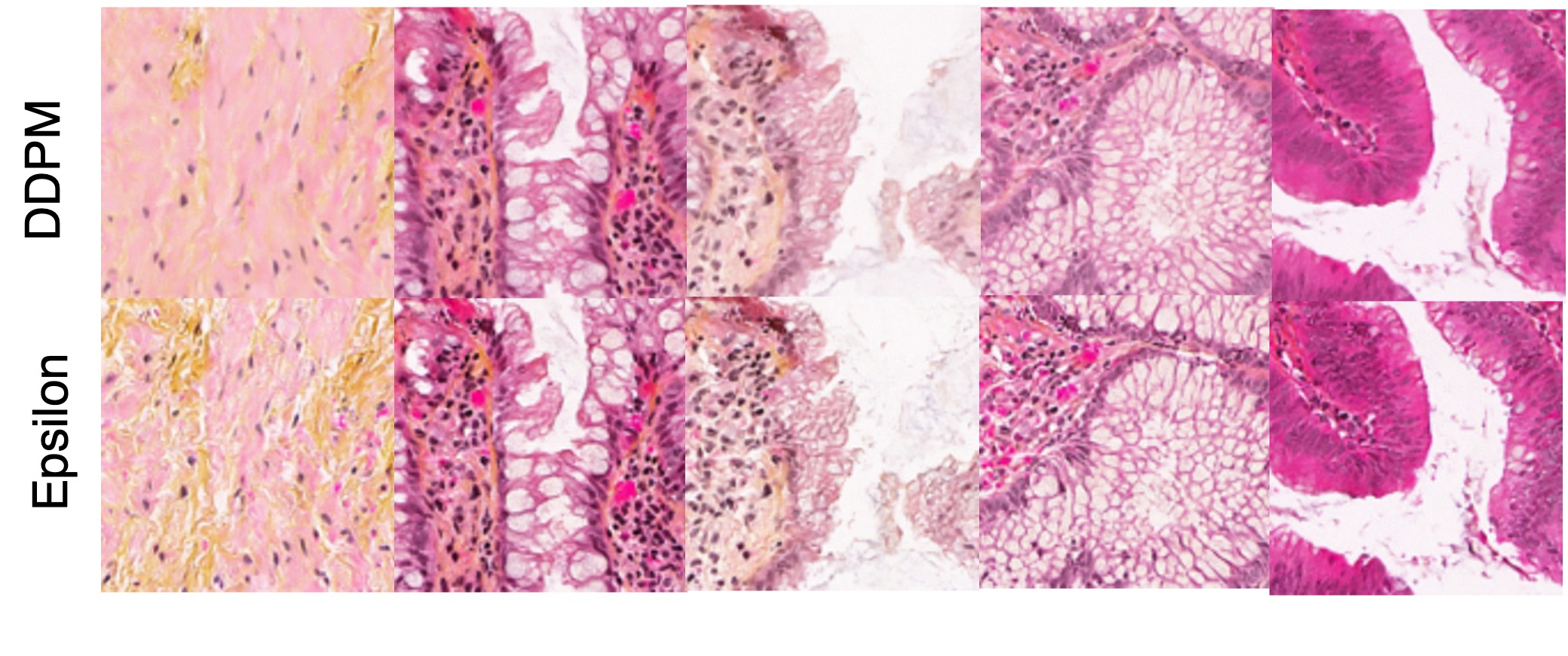}
\caption{Comparison of generated pathology images using DDPM (top row) and Epsilon sampling (bottom row).}
\label{fig:ddpm_epsilon_comparison}
\end{figure}

 Figure \ref{fig:ddpm_epsilon_comparison} shows a visual comparison of the results of the DDPM and Epsilon sampling methods applied to histopathology images. The upper row illustrates output from DDPM, while the lower row showcases Epsilon sampling results. Both methods can capture larger tissue structure from pathology images, but the DDPM images display more consistent structural details across different samples. In contrast, Epsilon sampling produces slightly varied textures, which could indicate greater flexibility in capturing diverse tissue characteristics. These observations suggest that while both methods are effective, DDPM might be more reliable for producing uniformly detailed images, whereas Epsilon sampling can capture a broader range of textural variations. This could be a reason for the higher FID score of the epsilon scaling method.

\subsection{Analysis on different patch size}
    \label{subsec:chapter4-patch-size-analysis}
  
      \begin{figure*}[h]
        \centering
        \includegraphics[width=1\linewidth]{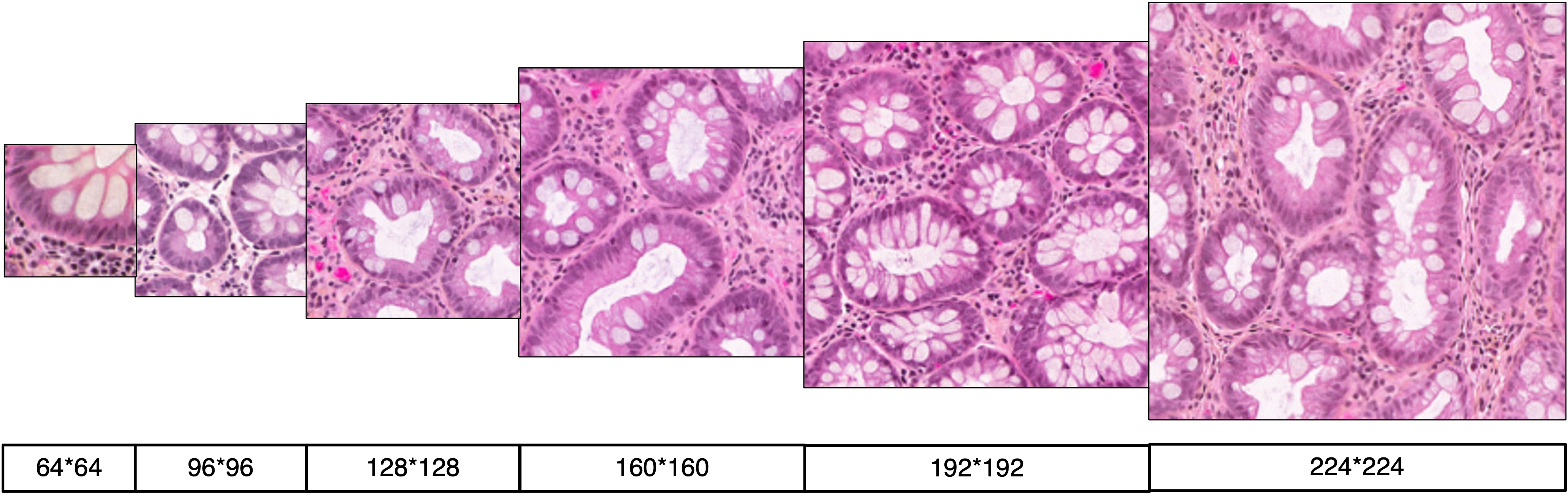}
        \caption{This series of images demonstrates the capability of our pre-trained model to generate histology slices at various resolutions, from 64x64 to 224x224. Each step up in resolution reveals more detail, illustrating the model’s ability to enhance image clarity from a single training setup.}
        \label{fig:image_resolution}
    \end{figure*}
    \begin{figure}[ht]
        \centering
        \includegraphics[width=1\linewidth]{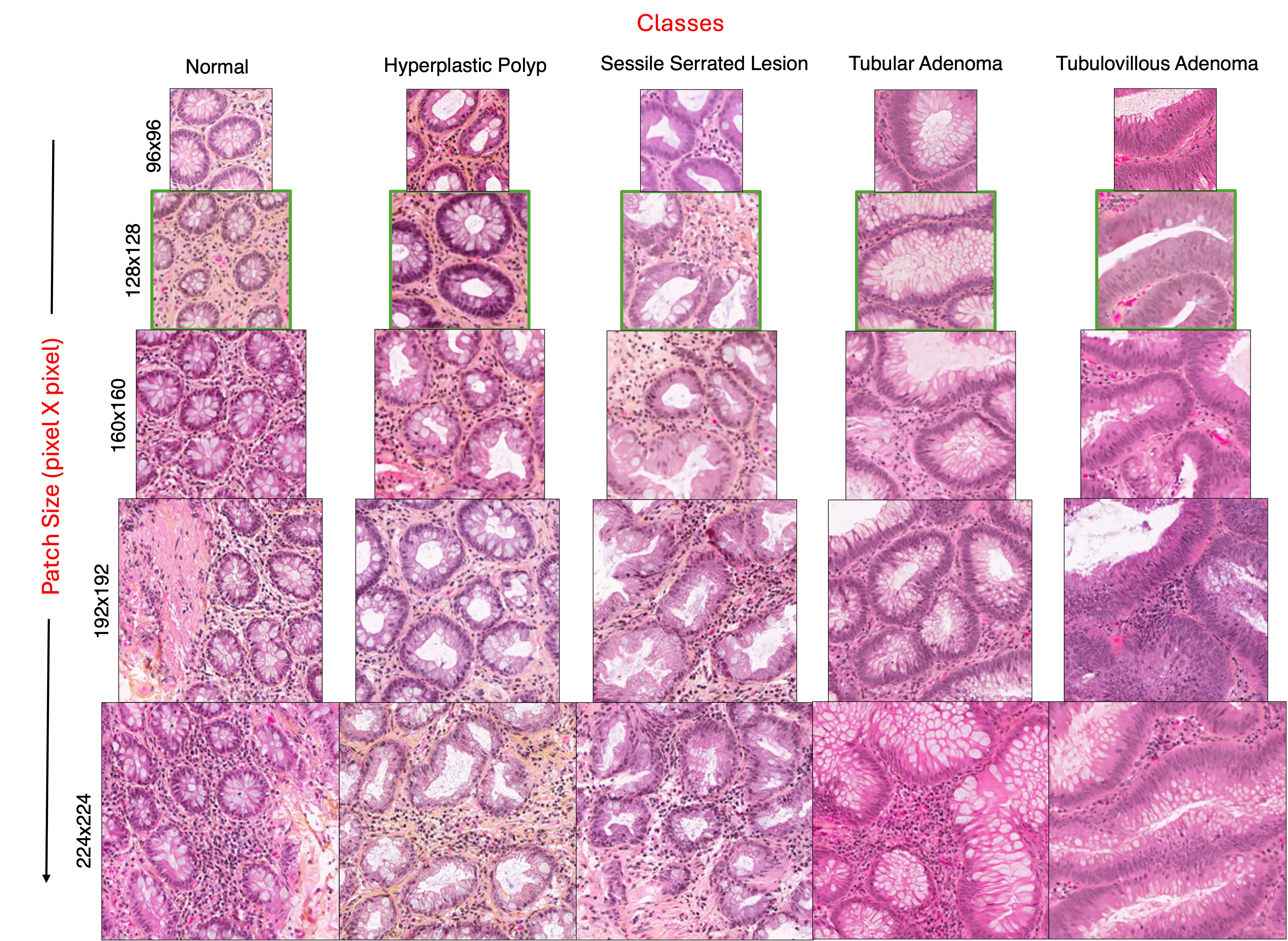}
        \caption{Patches generated for different patch size resembling different cell structure for five classes(128x128 is the training patch size used for diffusion model training)}
        \label{fig:fov_all}
    \end{figure}
    
    \begin{table}[ht]
    \centering
    \begin{tabular}{@{}l|l|l|l@{}}
    \hline
    \textbf{Input} & \textbf{FoV patch ($\mu m$)} & \textbf{Patch reshaped} & \textbf{FID}  \\ \hline
    Original Patch & 224 & 128 $\times$ 128 & \textbf{19.08}  \\ \hline
    Small Patch 
     & 112 & 64 $\times$ 64 & 161.01 \\
     & 168 & 96 $\times$ 96 & 33.71 \\ \hline
    Large Patch & 280 & 160 $\times$ 160 & 25.71 \\
     & 336 & 192 $\times$ 192 & 38.41 \\
     & 392 & 224 $\times$ 224 & 41.37 \\ \hline
    \end{tabular}
    \caption{Comparison of FID scores across different field of view and patch size.}
    \label{tab:fid_comparison}
    \end{table}
    
    The results indicate that the original patch size with a FOV of 224 and a patch resolution of 1.75 mpp achieved the lowest FID score of 19.08, reflecting the highest fidelity in generated images, which is expected as the images were trained on this model. However, examining the FID score for other patch sizes, especially those not used during training, provides insights into the model's ability to generate detailed images from different patch sizes.

     \begin{figure}[ht]
            \centering
            \includegraphics[width=0.6\linewidth]{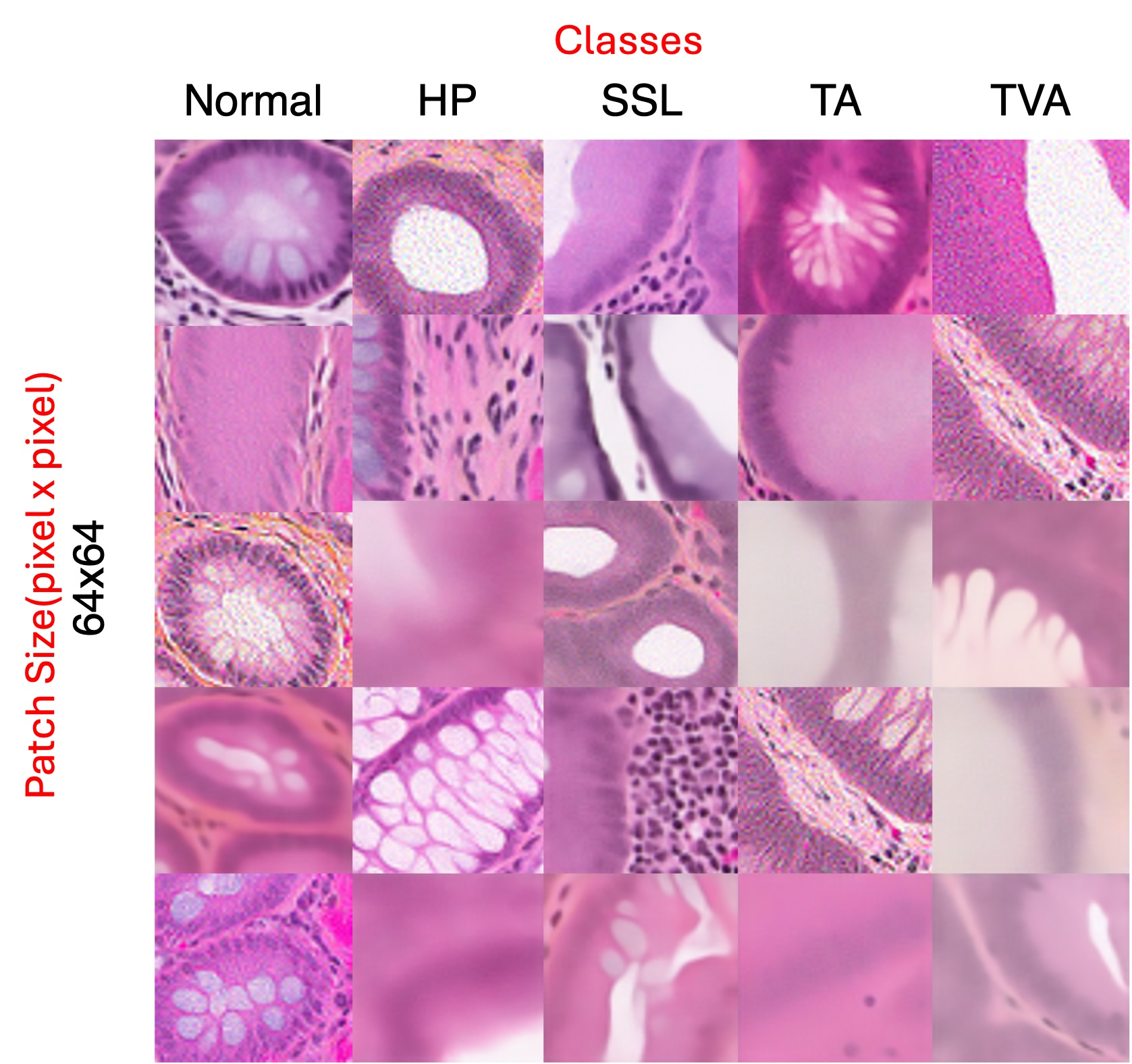}
            \caption{Patches generated for Patch size(64x64) which has the highest FID score}
            \label{fig:fov_patch_size64}
        \end{figure}

    This model is trained with a patch size of 128, and we experimented with generating images using different patch sizes, both smaller and larger. The table highlights the variation in image fidelity across different patch sizes and FOVs in generated pathology images. Smaller patch sizes, such as 64x64 and 96x96, exhibit higher FID scores of 161.01 and 33.71 respectively, indicating lower image fidelity. As the patch size increases to 160x160 and 192x192, the FID scores improve to 25.71 and 38.41 respectively, suggesting better image quality but with some variability. The 224x224 patch size, while larger, shows a slight decrease in fidelity with an FID score of 41.37. The visual representations in Figure \ref{fig:fov_all} demonstrate that larger patch sizes generally capture more detailed and high-level cellular structures across various classes, enhancing the potential utility of synthetic data in diagnostic workflows. The observed high FID score for the 64x64 patch size can be attributed to the fact that images generated at this resolution are often blurred and lack sufficient detail as shown in Fig \ref{fig:fov_patch_size64}. At such a low resolution, the model struggles to accurately generate fine patterns and intricate structures that are critical in pathology images. This inability to capture detailed cellular patterns results in poor-quality images, leading to the significantly higher FID score. In contrast, the 128x128 patch size allows the model to generate clearer images with more recognizable features and patterns, which contributes to a lower FID score. The blurriness for Patch size 64x64 reflects the model's limitation in generating complex, high-fidelity images at this lower resolution.
    
    Interestingly, even without training on certain pixel levels, the model can still perform well and generate quality images, demonstrating its robustness and versatility. This suggests that the generative model has the capacity to generalize beyond its training data to some extent, making it a valuable tool in scenarios where training data are scarce or diverse. This experiment highlights the model's adaptability and versatility.

    \subsection{Evaluation on Synthetic Dataset}
    
    The table below summarizes the classification performance on the KGH dataset under different scenarios, providing a clear understanding of how the use of generated data impacts model accuracy.
    
    \begin{table}[h]
    \centering
    \begin{tabular}{c|c}
    \hline
    \textbf{Dataset} & \textbf{ACC (\%)} \\
    \hline
    Real Dataset (PKGH\_224) & 89.95 \\
    Generated Dataset (PKGH\_224) & 88.62 \\
     Real + Generated (PKGH\_224) & 90.75 \\
    \hline
    Real Dataset  (PKGH\_336) & 94.06 \\
    Generated Dataset (PKGH\_336) & 92.44\\
    Real + Generated  (PKGH\_336) & 90.76 \\
    \hline
    \end{tabular}
    \caption{Classification accuracy summary for PKGH\_224 and PKGH\_336: Classification accuracy scores of a ResNet-50 on synthetic data. Higher ACC proves the effectiveness of DGM-generated(DDPM) synthetic samples in capturing significant features.}
    \label{table:classification}
    \end{table}
    
    The results reveal notable differences in model accuracy between real and generated datasets for both PKGH\_224 and PKGH\_336. When trained on the real PKGH\_224 dataset, the ResNet-50 model achieved an accuracy of 89.95\%. However, training solely on the generated PKGH\_224 dataset resulted in a slight drop in accuracy to 88.62\%. Interestingly, augmenting the real dataset with generated data led to an improvement in test accuracy to 90.75\%, indicating the added value of synthetic data in enhancing model performance.
    
    For the PKGH\_336 dataset, the model reached an accuracy of 94.06\% on the real dataset, and 92.44\% when trained on the generated dataset alone. However, combining real and generated data slightly decreased the accuracy to 90.76\%. Despite this, the PKGH\_336 dataset consistently demonstrated high performance, underscoring the robustness of the data, especially when leveraging the generated samples.
    
    The variations in accuracy between PKGH\_224 and PKGH\_336 can be attributed to differences in the Field of View (FOV) of the datasets. The larger FOV in PKGH\_336 captures more contextual information and intricate details, which likely enhances the model's ability to learn and generalize from the data, contributing to its superior performance.
    
    These findings highlight the potential benefits of incorporating synthetic data to boost model performance, particularly when real data is scarce or challenging to obtain. The overall accuracy scores affirm the effectiveness of DGM-generated synthetic samples in capturing essential features and improving the classification task.
\section{Conclusion}

In conclusion, we successfully tackled key challenges in histopathology by demonstrating the practical applications of diffusion generative models (DGMs) in medical imaging, particularly in generating high-quality synthetic datasets. The study confirmed that DGMs effectively learn from different patch resolutions, with larger patches providing superior results. While DDPM and LDM showed comparable performance despite their architectural differences, DDPM excelled in pixel space, and LDM generated high-quality images with fewer steps. Using two datasets, we found that larger FOV values yielded better FID scores and higher classification accuracy. This research highlights the potential of DGMs to enhance the robustness and accuracy of deep learning models in computational pathology, setting the stage for future advancements in the field.

\section{Future Work}

Future work will explore the potential of diffusion models to identify and analyze biomarkers for various diseases \citep{echle2021deep}. This involves developing methodologies to enhance the accuracy and reliability of biomarker discovery, leveraging the advanced capabilities of diffusion models to generate new subtle and critical features in medical images that are indicative of specific biomarkers. Future research can implement multi-class labeling techniques to enhance the classification of histopathology images. This involves conditioning diffusion models on multi-class labels to achieve more precise image analysis and interpretation. The aim is to improve the accuracy and robustness of classifiers in medical imaging, leading to better diagnostic and analytical outcomes.

These research directions aim to expand the current understanding and application of diffusion models in medical imaging and biomarker discovery. They will contribute to developing more accurate and privacy-sensitive healthcare solutions.

\section*{Acknowledgments}
The data collected for this study is supported by Huron Digital Pathology and Ontario Molecular Pathology Research Network (OMPRN) funding grant. We thank Resources for Research Groups(RRG)--Digital Research Alliance Canada (DRAC) and Computer Science Cluster at Concordia University for providing essential computational resources. Funding for this research is supported by NSERC-Discovery Grant RGPIN/05378-2022 and Concordia University FRDP Grant.

\bibliographystyle{unsrt}
\bibliography{refs}

\end{document}